# Analysis for the Overwhelming Success of the Web Compared to Microcosm and Hyper-G Systems

Bryar Hassan[1], and Shko Qader[2]

[1]Kurdistan Institution for Strategic Studies and Scientific Research, Sulaimani, Iraq.
‡[2]Sulaimani Governorate, Sulaimani, Kurdistan Region, Iraq.

**Abstract.** After 1989, Microcosm, Hyper-G, and the Web were developed and released. Each of these hypertext systems had some strengths and weaknesses over each other and their architecture were relatively different from each other. Standing above its competitors, the Web has become the largest and most popular information system. This paper analyses the reasons that why the Web was the first successful hypermedia system against all its competitors by looking at the architecture and evaluation of the Web and open hypermedia systems, and then three reasons beyond this success with some lessons to learn. Currently, Semantic Web is a recent development of the Web to provide conceptual hypermedia. More important, the study of the Web with its impact on technical, social and cultural, and economics agendas is introduced as web science.
**Word count:** 1497

## 1  Introduction

There have been many significant developments over the centuries in hypermedia systems, but the Web was presumably the most successful and popular one. Before the Web was released in 1991, there were two other hypermedia systems: Hyper-G and Microcosm. Currently, the Web is considered as the most popular and used distributed hypermedia systems, whereas the others have all around, yet they are disappeared. This paper presents an early hypermedia systems overview firstly. Secondly, open hypermedia system and the Web through the lens of their architecture and evaluation. Thirdly, main causes of growing the Web. Fourthly, some lessons to learn from succeeding the Web. Finally, recent developments of the Web.

## 2  Early History of Hypertext

Hypertext has a fairly long history. The most influenced hypertext systems are shown in this section.

**Memex.** this was proposed by bush in 1945[4]. Memex was an electro-mechanical device used for organising information and knowledge. It is considered as the forefather of all subsequent hypertext systems.

**Xandru.** after Memex, Nelson launched the Xandru project as a more comprehensive hypertext system in the 1960s and as a revision of Memex[5]. This project as an ideal hypertext system had the compelling features of link integrity and automatic version management. Furthermore, Nelson had invented both hypertext and hypermedia terms[6]. The former deals with text which organised in non-linear format and connected by links. The latter is the hypertext's extension, which is combination of both hypertext and multimedia.

**Dexter.** this was a formal reference model for an open hypertext system which was developed between 1988-1990. It was used to design existing and future hypertext systems[7]. the aim of this reference model was for system comparison as well as for interchange development and interoperability standards.

## 3  Open Hypermedia and the Web

### 3.1  Architecture

**Microcosm.** Microcosm was initially designed for desktop-based hypertext system as a research project at Southampton University. Then, it was changed to distributed hypermedia system[1]. Microcosm had three layers: application, link service, and storage layers.

**Hyper-G.** it was the midway between Microcosm and the Web and it was developed by Graz University around 1989-1990[1]. Hyper-G was also another distributed hypermedia system, which was based on client-server model[9]. It utilised its own protocol (HG-CSP) and its resource format (HTF).

**The Web.** it was initially proposed by Tim Berners-Lee at CERN to provide a distributed hypertext environment[3]. The Web architecture encompassed three essential technologies. Firstly, URI as an identifier was to address any resources on the Web. Secondly, network protocol such as HTTP, which defined how to receive and send messages between clients and servers. Finally, mark-up language such as HMTL was to specify resource format for documents.

### 3.2  Evaluation

The Web and open hypermedia systems had some strengths and weaknesses. Table-1 summarises the Web and open hypermedia systems' evaluation.

**Linking.** linking model of the Web was different from Hyper-G and Microcosm. The Web had simple node link model[13]. Nodes were interconnected with point-to-point, uni-directional, non-contextual, and no-typed links were used to present the Web's content. This simplicity of the today's Web leads to link dangling and broken. For example, an "Error 404" will be shown if the requested link is broken or not found. This simplicity of linking model, on the other hand, gives strength to the Web to be implemented easily[14]. Conversely, links were separated from nodes in open hypermedia systems[13]. This separation of data and links allowed user to navigate in various ways. Links were stored in a database in Microcosm. Dynamic linking was supported by Microcosm via generic links so that link destinations were managed on the fly automatically by the system. Likewise, Hyper-G used central link database to separate links from the nodes, but it was not as comprehensive as Microcosm. Consequently, it did not need to maintain the links manually in Microcosm and Hyper-G in case of broken. That means, their linking management was costless. This strength of linking model in Microcosm and Hyper-G, therefore, was also their weakness[13]. Hyper-G and Microcosm were complicated to implement technically peculiarly owing to their linking management.

**Scalability.** scalability is always the vital feature of hypermedia systems which includes performance with number of users[13]. The Web had generally more scalable than Microcosm and Hyper-G[16]. Firstly, the Web and Hyper-G were developed as large-scale hypermedia over the internet, whereas Microcosm was initially based upon intranet and it was also designed for cooperative and group activities. Secondly, the Web was designed as a decentralised system, whereas open hypermedia systems are centralised ones. Thirdly, open hypermedia had many built-in features: harmony browser in Hyper-G, dynamic linking in Microcosm, but the Web defined the minimum protocol nonetheless. Finally, Hyper-G document format was HFT and it had search engine facility, while the Web text format was HTML and it did not have search engine. Instead, it allowed third search engine[17]. However, Microcosm did not have any document format[16].

| Features | The Web | Hyper-G | Microcosm |
|---|---|---|---|
| Linking | Simple links and local anchor | All links except dynamic link | Supports all links |
| Simplicity | Yes | No | No |
| Scalability | Yes | No | No |
| Openness | Yes(URI) | No | No |
| Document format | Yes | Yes | No |
| Mark-up language | HTML | HTF | No |
| Proprietary | Non-proprietary | Proprietary | Proprietary |

**Table 1.** the Web and open hypermedia systems - evaluation

**Openness.** URI was used in the Web to identify any object via a simple text string, whereas Hyper-G and Microcosm did not have this concept[16]. This brought openness for the Web. Alternatively, There were document system in Hyper-G and Microcosm.

## 4    Growth of the Web

There were three main causes as shown in this section for growing the Web against its competitors.

**Technical.** there were six main technical causes of succeeding the Web. Firstly, simplicity and ease to use were the power of the Web, especially after developing and spreading web browser technology[16]. There is no doubt that the Web was not complicated compared to Hyper-G and Microcosm. Secondly, the Web had more flexibility than other hypermedia systems. For example, users were easily able to create plug-in such as search engines, and bi-directional links[15]. Thirdly, the Web was the most scalable hypermedia system in compare to Hyper-G and Microcosm. Fourthly, the Web was universally standardised and open protocols were provided as well[3]. Fifthly, the implementation of the Web was technically easy compared to the others thanks to linking model, protocols, and standardisation[13]. Lastly, the Web was compliant with all the Halasz's seven issues on hypertext[8], while the Hyper-G and Microcosm were not utterly compliant with it.

**Economical.** first and foremost, the Web was entirely non-proprietary[11], while Hyper-G and Microcosm were commercial products[2]. That means, either the Web would be used by everyone or it would not be so. Accordingly, users always prefer the Web rather than the other systems. Second, the Web was developed by CERN[3], whereas Microcosm and Hyper-G were developed by Universities[1]. Hence, this organisation was probably better for funding projects than Universities. Third, linking model and technical complications of Microcosm and Hyper-G were made difficulty to implement them economically[13].

**Social.** Hyper-G and Microcosm were developed in the Europe and attempted to spread thereof[16], while the Web was spread not only through the Europe but also through the USA on the Internet[3].Resultantly, both the Europe and USA, especially the latter one had probably more influence on people globally than only the Europe.

## 5    Lessons to Learn

Overwhelming success of the Web addresses three vital lessons. First, the provision of document format, open protocols and universal standards were the cornerstone of its success. Second, the Web was initially showed the idea of "scruffy works"[16]. This creates opportunity to be more simpler and usable by users easily. Final, freeness, openness, decentralisation, and easy to use of the Web were the vital key of using it by users.

## 6   Recent Developments

At the present time, Semantic Web is considered as the next stage of the Web development [11]. The aim of Semantic Web is to change the current machine-readable web into machine-understandable. Furthermore, COHSE endeavours to bring conceptual hypermedia into the Web in order for the Web to be able to implement dynamic linking as Microcosm did so[12]. Conceptual hypermedia is presumably considered as Semantic Web and COHSE can be counted as its application. Meanwhile, technical, social and cultural, and economics agendas have all effect on shaping the future of the Web. An interconnection between those perspectives is needed to form the nature of the Web in the future. Therefore, a new discipline named web science is such an important field[10][16]. Based on that, web science discipline has initiated recently by Southampton University and MIT so as to analyse and monitor these developments of the Web.

## 7   Conclusion

This paper has introduced early hypermedia systems, architecture and evaluation of open hypermedia systems and the Web, technical, economical, and social reasons for becoming the Web most popular against its competitors with three lessons to learn from this success, and recent development of the Web. Each of these hypermedia system had some advantages and disadvantages over each other. Incidentally, technical, economical, and social agendas and interconnections between them have impact on playing the role of forming and anticipating the future of the Web. Accordingly, web science as a new discipline has initiated recently. In respect to more recent developments of the Web, conceptual hypermedia or Semantic Web as a description of the Web has introduced new trends for hypermedia and the Web.